\documentclass[10pt, conference,compsocconf]{IEEEtran1.7a}

\usepackage[T1]{fontenc}

\usepackage{cite}
\usepackage{amsmath,amssymb,amsfonts}
\usepackage{algorithmic}
\usepackage{graphicx}
\usepackage{textcomp}
\usepackage{xcolor}
\usepackage[inline]{enumitem} 
\usepackage{enumerate} 
\usepackage{multirow}
\usepackage{balance}
\usepackage{hyperref}
\usepackage{slashbox} 
\usepackage[ruled,vlined,linesnumbered]{algorithm2e}

\usepackage{tablefootnote}

\SetCommentSty{mycommfont}

\usepackage{pifont}
\newcommand{\cmark}{\ding{52}}%
\newcommand{\xmark}{\ding{54}}%

\def\BibTeX{{\rm B\kern-.05em{\sc i\kern-.025em b}\kern-.08em
    T\kern-.1667em\lower.7ex\hbox{E}\kern-.125emX}}

\renewcommand{\IEEEbibitemsep}{0pt plus 2pt}
\makeatletter
\IEEEtriggercmd{\reset@font\normalfont\footnotesize}
\makeatother
\IEEEtriggeratref{1}

\begin{document}
\bstctlcite{IEEEexample:BSTcontrol}


\title{SECDA: Efficient Hardware/Software Co-Design of FPGA-based \\ DNN Accelerators for Edge Inference}


\author{\IEEEauthorblockN{Jude Haris\IEEEauthorrefmark{1},
Perry Gibson\IEEEauthorrefmark{1},
Jos\'e Cano\IEEEauthorrefmark{1},
Nicolas Bohm Agostini\IEEEauthorrefmark{2},
David Kaeli\IEEEauthorrefmark{2}},
\IEEEauthorblockA{
        \begin{tabular}{cc}
            \IEEEauthorrefmark{1}{University of Glasgow, UK} 
            \IEEEauthorrefmark{2}{Northeastern University, USA}
        \end{tabular}
  }
}


\maketitle


\begin{abstract}

Edge computing devices inherently face tight resource constraints, which is especially apparent when deploying Deep Neural Networks (DNN) with high memory and compute demands. FPGAs are commonly available in edge devices. Since these reconfigurable circuits can achieve higher throughput and lower power consumption than general purpose processors, they are especially well-suited for DNN acceleration. However, existing solutions for designing FPGA-based DNN accelerators for edge devices come with high development overheads, given the cost of repeated FPGA synthesis passes, reimplementation in a Hardware Description Language (HDL) of the simulated design, and accelerator system integration.

In this paper we propose SECDA, a new hardware/software co-design methodology to reduce design time of optimized DNN inference accelerators on edge devices with FPGAs. SECDA combines cost-effective SystemC simulation with hardware execution, streamlining design space exploration and the development process via reduced design evaluation time. As a case study, we use SECDA to efficiently develop two different DNN accelerator designs on a PYNQ-Z1 board, a platform that includes an edge FPGA. We quickly and iteratively explore the system's hardware/software stack, while identifying and mitigating performance bottlenecks. We evaluate the two accelerator designs with four common DNN models, achieving an average performance speedup across models of up to 3.5$\times$ with a 2.9$\times$ reduction in energy consumption over CPU-only inference. Our code is available at \texttt{https://github.com/gicLAB/SECDA}

\end{abstract}



\section{Introduction}

Deep Neural Networks (DNNs) have demonstrated high accuracy in learning tasks, such as image classification~\cite{alex2012}, speech recognition~\cite{Zhang2016TowardsES} and many more. However, current solutions that attempt to deploy DNNs on low-power and resource-constrained edge devices (e.g., smartphones, tablets and wearables) are inefficient~\cite{Lane2015}, presenting challenges that can span several levels of the hardware/software stack to run efficiently~\cite{iiswc_2018}. 
To address these inefficiencies, hardware-based optimizations have been proposed to reduce DNN inference costs. This active research area includes ISA-level extensions to CPUs and GPUs~\cite{Ottavi2020,Markidis2018}, as well as TPUs~\cite{jouppi2017} and other custom hardware solutions for FPGAs and ASICs~\cite{chen2019,kwon2018}.

There are a variety of tools available for developing FPGA-based DNN accelerators for edge devices~\cite{alwani2016,liu2017}, some of which feature model-specific tuning~\cite{umuroglu2017,moreau2019a,zhang2018a}.
Flexibility is valuable, since the layers and sizes of DNNs can vary, especially given the rapid introduction of new and novel DNN architectures.   
A good example of the range of DNNs is the large convolutional layers of InceptionV3~\cite{Szegedy_2016_CVPR}, as compared to the small depth-wise separable convolutions of MobileNets~\cite{howard2017}.

However, the process of developing DNN accelerators using these tools is generally poorly documented, and prior work focuses on their features and results, rather than their design methodologies~\cite{qin2020,lu2017}.
Since resources are more limited on edge FPGAs, and DNNs workloads are large in terms of their memory footprint and computational demands, a given DNN is unlikely to fully fit on an accelerator.
Thus for DNN inference, the accelerator must operate in close communication with the CPU, which requires careful co-design with the host CPU code to ensure that data is managed efficiently.
Therefore, an effective design methodology for a DNN accelerator design should, for a given set of hardware resource constraints, produce performant accelerators that effectively leverage available resources, and can respond quickly to changing workload requirements (e.g., introduction of new types of layers or operations).

The process of mapping candidate hardware designs to an FPGA, known as synthesis~\cite{Reese2006}, is a time-consuming process that can take minutes to hours, depending on the complexity of the design. Compounding this with the number of iterations in a typical design process, synthesis can create a clear bottleneck in the hardware development process. Existing solutions either accept the synthesis costs~\cite{liu2011}, surrender low-level design fidelity~\cite{openCL}, or develop accelerators using a purely simulation-based approach~\cite{stonne-iiswc2021,xi2019} that frequently results in non-synthesizable hardware solutions.

Motivated by the challenges of making informed accelerator design choices while efficiently evaluating them, we propose SECDA (\emph{SystemC Enabled Co-design of DNN Accelerators}), a new hardware/software co-design methodology to efficiently produce optimized DNN inference accelerators for edge devices using FPGAs. 
SECDA excels in the following five key features:
\begin{enumerate*}[label=(\roman*)]
    \item \label{itm:data-design} \emph{Design Control}: ability to easily specify low-level behavior of accelerator designs;
    
    \item \label{itm:e2e} \emph{End-to-end Evaluation}: ease of performing full DNN models inference with candidate designs;
    
    \item \label{itm:codesign} \emph{Driver Co-Design}: the CPU-side software driver is developed in tandem with the hardware accelerator, so that design trade-offs can be considered early;

    \item \label{itm:itegration} \emph{System Integration}: candidate accelerator designs are easily realized on FPGA hardware; and

    \item \label{itm:sim_speed} \emph{Simulation Speed}: simulation of candidate accelerator designs is faster than synthesis, and provides sufficient performance information to guide design improvements.
\end{enumerate*}
SECDA uses SystemC~\cite{IEEE2012systemc} as an accelerator simulation framework, which also allows candidate designs to be iterated upon efficiently. 
Using SystemC High Level Synthesis (HLS), we can produce a synthesizable design from the same accelerator definition. 
We leverage SystemC's modularity to reduce the time required to explore design changes, by reusing and adapting existing components. 

The co-design of a software accelerator driver and a hardware accelerator is achieved by integrating SystemC's simulation features within the target edge-based DNN framework, allowing designers to quickly test potential optimizations such as varying data transfer and tiling strategies. 
Embedding the simulation environment and the hardware accelerator into the same software environment reduces the costs of exploring hardware/software co-design trade-offs via simulation, relative to synthesizing the design on an FPGA with every change.

We demonstrate the utility of SECDA with a case study that targets acceleration of General Matrix Multiplication (GEMM), a  heavily used kernel in convolutional layers, and the most computationally-expensive portion of many DNNs~\cite{10.1145/3184407.3184423}. For this case study, we develop two accelerator designs, a Vector MAC and a Systolic Array. The DNN framework used is TensorFlow Lite (TFLite), a mobile-friendly version of TensorFlow~\cite{abadi2016}. The target device is the PNYQ Z1 board~\cite{pynqz1}, a platform with a dual-core CPU and an edge FPGA. 

The contributions of this paper include the following:

\begin{itemize}
    \item We motivate the need for a better design methodology for DNN accelerators, and define five key features that future methodologies should support, which are essential for fast design exploration and integration of hardware accelerators for DNNs using FPGAs.

    \item We propose SECDA, a new design methodology to efficiently explore the design space of DNN accelerators for edge devices with FPGAs, and quickly arrive to optimized solutions. We show how SECDA excels in the five key features and reduces the time to obtain efficient designs.
    
    \item We demonstrate the capabilities of SECDA by efficiently designing two GEMM accelerator designs for DNNs using the PYNQ-Z1 platform~\cite{pynqz1}, a Systolic Array (SA) architecture and a Vector MAC (VM) architecture.
   
    \item We evaluate the two accelerator designs with four common DNN models and achieve an average performance speedup across models of up to $3.5\times$ with a $2.9\times$ reduction in energy consumption compared to a CPU-only baseline.
\end{itemize}

\section{Motivation}
\label{sec:motivation}

We separate the development of hardware accelerators for DNNs into two stages: i) the design of hardware primitives, composing them to construct an accelerator architecture and defining the software stack to support the accelerator design; ii) an optional secondary stage where the designer exposes aspects of the design as a set of templates with tunable parameters (e.g., buffer sizes, number of processing elements) and allows automated design space exploration tools to find more performant designs.
We refer to this latter stage as \emph{accelerator frameworks}, which include VTA~\cite{moreau2019a}, DNNBuilder~\cite{zhang2018a}, and FINN~\cite{umuroglu2017}.
This optional stage of design is valuable, especially when the goal is to optimize DNN hardware accelerator designs for a range of specific DNN models.

The first stage, which we name as \emph{accelerator design methodologies}, is the focus of this paper. 
We make the distinction between the two stages to highlight that though there is a rich and growing literature for the second stage, for the first stage there is a lack of discussion about the design methodologies, and how they can better accommodate the features of DNNs to address inefficiencies with the current process of creating edge based accelerator solutions. This is the motivation for defining our new methodology SECDA.

In Section~\ref{subsec:desire_feats} we elaborate on \emph{accelerator design methodologies} and describe five key features they should fully support to be efficient. 
Then in Section~\ref{subsec:method_comp} we discuss and compare common FPGA design methodologies in regards to these five key features.


\subsection{Key features of DNN Accelerator Design Methodologies}
\label{subsec:desire_feats}

There are two important characteristics of the workload to acknowledge when designing hardware accelerators for DNNs using edge FPGAs.
First, though DNNs are getting more efficient~\cite{hernandez2020}, they are still large programs with ever-increasing memory and compute demands. Many DNNs involve a large number of parameters and operations, making them difficult to fit on a typical edge FPGA without partitioning them into stages.
Secondly, DNNs feature a variety of operations (in terms of layers), with varying frequency and costs. It may be preferable to fallback to the CPU for less frequent operations and focus accelerator resources on the most expensive layer types.
Both of these characteristics mean that DNN accelerator designs must be closely designed with the CPU host-side software to ensure efficient balancing of the workload.

To effectively tackle these two characteristics, we define five key features that accelerator design methodologies are required to excel in, to efficiently produce optimized FPGA-based DNN accelerators:
\begin{itemize}
    \item \emph{Design Control}: The degree of control given to the designer for both high-level and low-level features, such as the overall dataflow at a high-level and behavior and interconnection of individual components at a low-level, balancing model depth against overall simplicity.
        
    \item \emph{End-to-end Evaluation}: Inference evaluation (either in simulation or in hardware) of full DNN models is key. 
    This process should be as fast as possible to keep design iterations short. 
    Benchmarking only single layers may cause the designer to miss the bottlenecks that only emerge with realistic workloads.
    
    \item \emph{Driver Co-Design}: The interface between the accelerator and the DNN framework can play a pivotal role in the efficiency and performance of the design~\cite{wang2019,xi2019}. A good design methodology should enable the designer to co-design the software driver and the hardware accelerator, thus allowing to explore different degrees of workload offloading, and creating an effective workload balance between the CPU and the accelerator. 

    \item \emph{System Integration}: 
    We need both ease and speed in the process of mapping a proposed accelerator design to an FPGA. 
    This includes the integration of the accelerator with the DNN framework software. The goal is to have minimal overhead in realizing a design on real hardware. 

    \item \emph{Simulation Speed}: Leveraging simulation can reduce the time taken for logic synthesis within the design process. Hence, it is crucial that simulation is fast, accurate, and does not become the bottleneck of the design loop.
\end{itemize}


\subsection{Comparison of Methodologies}
\label{subsec:method_comp}

Table~\ref{tab:features} compares different state-of-the-art design methodologies based on the previous five key features against SECDA. Next, we discuss the details of each methodology.

\textbf{OpenCL} \cite{openCL} uses a host-device programming model, where the host code (i.e., the driver) prepares and transfers data to be executed by the device (i.e., the accelerator). Hence, the approach allows for the co-design of the driver. The device code is written in high-level OpenCL code which defines computation kernels that perform the processing of the target workload. This high-level code is translated into a synthesizable hardware design. The designer defines the computation kernels to be accelerated, being able to configure the number of hardware instances each kernel is allocated. The higher the number of instances, the greater number of instructions executed in parallel. The level of \emph{design control} offered by OpenCL can be restrictive, since the designer cannot easily define the low-level behavior of the accelerator, e.g. at the transaction level, which can enable control of each sub-component and interfaces between them. The Intel FPGA SDK for OpenCL\cite{AlteraOpenCL} allows emulation of accelerator designs on x86 machines, which allows for verification of the behavior of the design. To gather dynamic performance of the hardware accelerator, the designer has to perform slow cycle-accurate simulation, or hardware profiling on the target FPGA, which can take hours.

\begin{table}[t]
\caption{\label{tab:features}Comparison of methodologies with five key features.}
\centering
\fontsize{6.5}{9}\selectfont
\begin{tabular}{|c||c|c|c|c|c|}
\hline
\textbf{Feature \textbackslash \: Methodology} & OpenCL & HDL & SMAUG & SECDA \\ \hline 
Design Control & Low & Very High & Medium & High \\ \hline
End-to-End Evaluation & \cmark & \xmark & \cmark & \cmark \\ \hline
Driver Co-Design & \cmark & \xmark & \cmark & \cmark \\ \hline
System Integration & Simple & Difficult & N/A & Simple \\ \hline
Simulation & Slow & Very Slow & Slow & Fast \\ \hline
\end{tabular}
\end{table}

\textbf{Hardware Description Language} (HDL) based design flows use highly detailed hardware descriptions in languages such as Verilog~\cite{verilog} and VHDL~\cite{vhdl}, to define the desired behavior of the accelerator. While this approach allows for fine-grained hardware designs, it comes with high development time costs, resulting in high code-base complexity and strict size requirements to define a design~\cite{Pelcat2016}, as compared to HLS or OpenCL-based solutions. Additionally, although HDL solutions can use RTL simulator to provide cycle-accurate simulation, the level of simulation detail makes the process much slower than non-RTL based simulations. An HDL-based approach to designing accelerators does not lend itself well to co-design of the host driver, or end-to-end evaluation, since RTL simulators are testbench-based and inherently slow.

\textbf{SMAUG}~\cite{xi2019} provides a simulation-based design methodology that uses gem5-Aladdin~\cite{gem5-aladdin} to perform full system simulation of the host system, the off-chip memory accesses and the accelerator design itself. While this approach provides high fidelity in terms of design performance insights, the simulation speed is very slow due to simulation of the entire system (e.g., several hours for ResNet50). Rather than integrating with an existing DNN framework, models must be redefined using SMAUG's Python API. SMAUG does not offer an approach where a design can be directly synthesized to a target FPGA and integrated with a DNN framework of choice.

\textbf{SECDA} uses the SystemC programming model to define the behavior of an accelerator at a transaction-level, and uses HLS to produce synthesizable designs. This provides a high degree of design control, while mitigating the issues of cumbersome HDLs. SECDA integrates a SystemC simulation environment within the target DNN framework, allowing co-design of the accelerator driver, and simulation of end-to-end inference. Unlike SMAUG, SECDA does not simulate the full host system, avoiding the large overheads and keeping simulation times in the order of minutes. 
For most design iterations, we argue that full host system information is not relevant, since most design choices are related to the performance within the accelerator.
Once the accelerator designs are refined through simulation, we identify issues related to the full system, such as off-chip memory accesses. We leverage SECDA's HLS capabilities to test on real FPGA hardware. Thus, avoiding the high simulation costs seen in methodologies such as SMAUG, and expensive hardware synthesis with each design iteration. For rapid design of DNN accelerators for edge FPGAs, SECDA achieves a good trade-off in terms of simulation fidelity, design granularity, and ease of deployment on real hardware.


We now compare the development time of these methodologies with illustrative estimates of the ``idle'' time spent waiting for evaluation of candidate designs. For SECDA, we compute this time ($E_{\mathrm{t}}$) with the following equation:

\begin{equation}\label{eqn:secda_time}
        E_{\mathrm{t}} = \mathrm{\#Sim} * (C_{\mathrm{t}} + \mathit{IS}_{\mathrm{t}}) + \mathrm{\#Synth} * (S_{\mathrm{t}} + I_{\mathrm{t}})
\end{equation}

Where $\#Sim$ is the number of simulated design iterations performed; $C_{\mathrm{t}}$ and $IS_{\mathrm{t}}$ are the times to compile and run an end-to-end inference in simulation, respectively; $\#Synth$ is the number of hardware synthesis passes performed; $S_{\mathrm{t}}$ is the time to perform logic synthesis of the accelerator design; and $I_{\mathrm{t}}$ is the time to perform inference on the FPGA. 
Since the time for $S_{\mathrm{t}}$ dominates the overall time, minimizing the number of logic synthesis performed is desirable.

For OpenCL and HDL, their simulation costs are significantly higher than SECDA since they use cycle-accurate simulations. 
We could also follow a design methodology which eliminates simulation, and only relies on iterations using logic synthesis alone. The equivalent time spent waiting for evaluation results is given by: 

\begin{equation}\label{eqn:alt_time}
        E_{\mathrm{t}} = (\mathrm{\#Sim} + \mathrm{\#Synth}) * (S_{\mathrm{t}} + I_{\mathrm{t}})
\end{equation}

Finally, a design methodology using full system simulation to perform all design iterations (e.g., SMAUG), would have a similar idle evaluation time estimate as in Equation~\ref{eqn:alt_time}, but with the simulation time replacing the synthesis time.

\begin{equation}\label{eqn:alt_sim_time}
        E_{\mathrm{t}} = (\mathrm{\#Sim} + \mathrm{\#Synth}) * (C_{\mathrm{t}} + \mathit{IS}_{\mathrm{t}})
\end{equation}

However, the simulation cost $I_{\mathrm{t}}$ would be significantly higher than SECDA's low-cost SystemC simulation due to a more complex simulation, which we argue is not needed in SECDA due to our two-stage approach.
Both OpenCL and HDL-based methodologies can use either, or a mix of, the approaches described by Equations~\ref{eqn:alt_time} and \ref{eqn:alt_sim_time}.
However, for these types of methodologies, both simulation and synthesis are expensive in terms of development time, whereas in SECDA we take advantage of fast simulation, which is sufficient for most design iterations, and only occasional synthesis.

\section{SECDA Methodology}
\label{sec:design}

In this work, we propose \emph{SECDA} (\emph{SystemC Enabled Co-design of DNN Accelerators}), a new hardware/software co-design methodology to efficiently design optimized DNN inference accelerators for edge devices with FPGAs. 
SECDA provides fast accelerator design space exploration, integrating software and hardware design choices, and reducing barriers when evaluating designs in real edge hardware. 
As discussed in Section~\ref{sec:motivation}, SECDA targets DNN inference at the edge, since edge accelerators are inherently more resource constrained.
Based on the characteristics of DNN workloads, SECDA focuses heavily on efficient host-accelerator communication, which requires careful co-design to ensure that data is managed efficiently. 

\begin{figure}[!t]
 \centering
 \includegraphics[width=0.9\columnwidth]{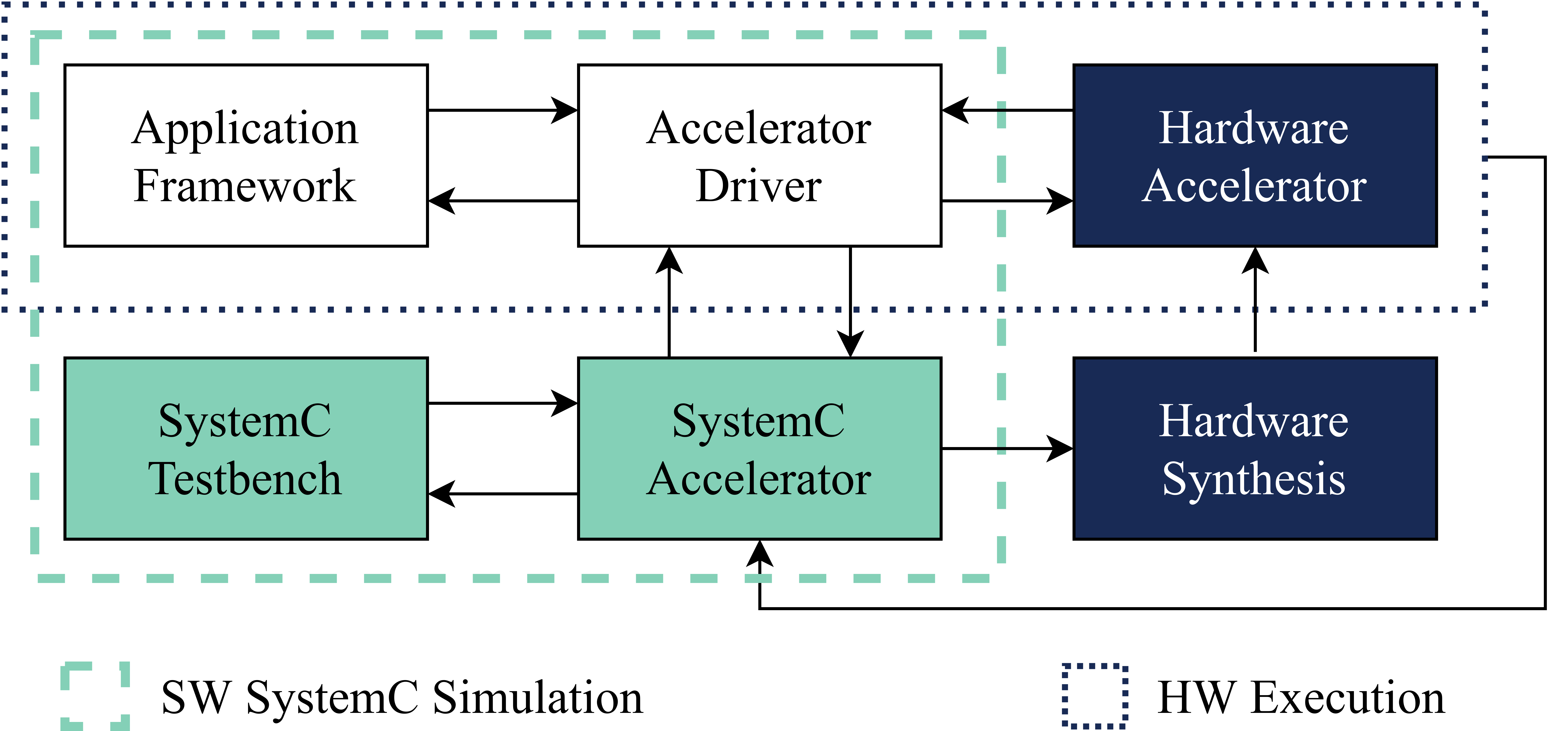}
 \caption{\label{fig:overview} Overview of the SECDA methodology. Components in the dashed lines correspond to the design in simulation and components in the dotted lines correspond to the design running on real hardware. \emph{Application Framework} and \emph{Accelerator Driver} software are common to both.}
\end{figure}

Figure~\ref{fig:overview} shows a high-level overview of the proposed methodology. The following sections provide details on the key components of the methodology, including how the components are interconnected to form the SECDA design loop.


\subsection{Application Framework}

The \emph{Application Framework} is the DNN software framework which runs the target DNN models, from which we offload work to the accelerator.
We characterize it as software that could run the full workload independently of the accelerator. 
For example, edge inference specific versions of popular deep learning frameworks such as TensorFlow's~\cite{abadi2016} TFLite and PyTorch~\cite{paszke2017automatic} Mobile, which reduce the feature set of the frameworks to run inference more efficiently using fewer resources.
In SECDA, we ensure that the \emph{Application Framework} is integrated early in the design cycle so that the accelerator development is informed by real workloads, and that co-verification is improved by avoiding software compatibility issues, such as misaligned data or conflicting data types.
To ensure that the designer has a realistic understanding of the bottlenecks in their designs, the SECDA methodology is instantiated with support for running full DNN workloads from the start, so that they can focus on the most relevant aspects of their design for the target workloads.

\subsection{Accelerator Driver}
\label{subsec:acc_driver}

The \emph{Accelerator Driver} is the software component in the co-design methodology, the bridge between the \emph{Application Framework} and the hardware accelerator. It is responsible for managing aspects such as data preparation, output data unpacking, control flow and memory management for DMAs, and thread synchronization. The efficiency of the \emph{Accelerator Driver} can be very impactful on the overall runtime performance~\cite{wang2019,xi2019}, hence why driver co-design is a key feature of SECDA. For example, the design of the input data preparation stage is crucial, because the data format of the \emph{Application Framework} may not be suited for a given accelerator design. Non-accelerated CPU-code may reshape data to leverage vector instructions, but we may prefer to reshape data differently to leverage the design of a given accelerator's architecture.
Thus, we may face co-design trade-offs where we must choose a data format which balances the efficiency of processing it on our hardware design while also enabling efficient CPU-side driver conversion to-and-from this format.

With both input preparation and output unpacking stages, the driver should ensure that data transfers between the accelerator and main memory are performed efficiently, since they can dominate both inference time and energy costs for DNN accelerators~\cite{sze2017}.
The driver is also responsible for the workload balance between accelerator and CPU, and should ensure that the aforementioned stages are pipelined such that the CPU is not idle while the accelerator is working.

We co-design the \emph{Accelerator Driver}, along with the accelerator, in an end-to-end \emph{SystemC Simulation} environment integrated with the \emph{Application Framework}. As observed in Figure~\ref{fig:overview}, the \emph{Accelerator Driver} is reused in both simulation and hardware evaluation, the latter giving performance data such as data communication latency between host and accelerator, on system components not modeled in detail by the cheap simulation such as off-chip memory accesses.


\subsection{SystemC Simulation}
\label{subsec:sysc_sim}

SystemC~\cite{IEEE2012systemc} is a C\texttt{++} library that models and simulates the behavior of hardware designs. We use SystemC for Transaction-Level Modeling (TLM)~\cite{Maillet-Contoz2005}, which simulates complex designs without the overhead of exact register-level details, while still ensuring bit-level accuracy. SystemC Simulation is the cornerstone of the SECDA methodology. Using simulation combined with HLS we can gain insight into candidate designs. SECDA is over an order of magnitude faster than using logic synthesis alone to configure the FPGA in our case study. In SECDA, we use two levels of SystemC Simulation (testbench and end-to-end) to further refine our co-designed steps, one for designing low-level components, and the other for evaluating the full accelerator design.

\textbf{SystemC Testbench simulation} is based on unit testing the accelerator design and its components on sets of input data, which enables developers to iteratively design accelerator components without running a full workload. Using SystemC HLS, we feed performance estimates such as clock cycle costs, and overall resource utilization for each component into the design simulation model. The testbench environment allows for quick development of designs without needing compatible drivers to interface with a full scale DNN framework.

\textbf{End-to-end SystemC Simulation} runs entire DNN models using our candidate accelerator designs, with integration of the \emph{Application Framework} via the \emph{Accelerator Driver}. This higher level of abstraction tests the correctness of the full system and leverages the accelerator's per-component performance estimates to show metrics for full-workloads. Using end-to-end simulation, we capture behavioral and performance information of the accelerator when simulated with the input data produced by any given model. 

The metrics captured from these simulations can include the number of total clock cycles spent within the accelerator, BRAM utilization, processing element utilization and various other metrics. These metrics can motivate further design iterations and highlight components representing bottlenecks. For example, we can identify inefficient processing elements or provide guidance on whether to perform wider exploration of the design space, investigate different data-flow strategies, or increase resource utilization. The accuracy we observe in terms of clock cycle count is over 99\% in our case study, when compared to the same designs synthesized on hardware. 
The simulation accuracy level achieved within the case study should be replicable in more complex designs, as SystemC timing models of low-level components are composable and hierarchical.


\subsection{Hardware Synthesis}

A key step of SECDA is mapping a candidate accelerator design to real hardware in order to collect data which the designer uses to improve the overall application performance (e.g., in terms of inference time and/or energy consumption). FPGAs are an ideal platform for testing hardware designs, as well as running workloads. When an accelerator design meets the performance targets in the simulation, we can map our SystemC Accelerator onto the FPGA using HLS, followed by logic synthesis. Then we can perform \emph{Hardware Evaluation} running the \emph{Application Framework} with the \emph{Hardware Accelerator} using the \emph{Accelerator Driver}, as shown in Figure~\ref{fig:overview}. This involves an end-to-end evaluation of target DNN models using the synthesized accelerator design.

Logic synthesis is one of the most costly stages of any FPGA-based design process. Hence, we opt to perform most of our accelerator design space exploration using SystemC Simulation. Compared to hardware synthesis, compiling the same design to run in SystemC Simulation is much faster, around $25\times$ faster for the Vector MAC design in our case study (Section \ref{sec:case_study}). The advantage of running the application on the FPGA synthesized accelerator is that we collect actual performance values, rather than the estimates generated through simulation. Following this methodology can highlight bottlenecks created by the host system, such as data transfer overheads which are not modeled in the low-cost simulation.


\subsection{SECDA Design Loop}

SECDA relies on two different iterative design loops to explore the edge accelerator design space for DNNs. The most frequently used design loop iterates through inexpensive SystemC simulations, and the second loop involves hardware benchmarking on edge FPGAs. Hardware benchmarking requires logic synthesis, which is very time consuming. Thus, SECDA aims to minimize the number of times this occurs. 

SECDA enables the designer to choose between the two iterative design loops. The SystemC simulation design loop is appropriate when profiling the performance of the individual components of the accelerator, or the overall performance of data processing within the accelerator. 
The performance profile of a given DNN model within the accelerator can also be evaluated in simulation, which with a diversity of models can highlight weaknesses in the hardware design.
The hardware benchmarking design loop is appropriate when the designer is interested in accurate performance data of DNN models; 
in particular the data transfer latencies between off-chip and on-chip memory, which is not modeled by simulation to limit the cost of simulation.

SECDA achieves increased productivity by giving the designer this choice. Through the low-cost SystemC simulation, the designer can effectively avoid expensive hardware synthesis until an efficient accelerator design is fully developed. At this point, the design can be mapped to the target FPGA for actual performance metrics to be gathered.

Full-system simulation, as used in SMAUG~\cite{xi2019}, would avoid the need to synthesize. However this simulation often takes longer than synthesis, thus we choose to utilize a less expensive simulation $+$ a hardware-evaluation approach within SECDA. Hardware synthesized designs are evaluated and used to inform further iterations in simulation, until the final design is chosen to meet the expected performance targets, such as reduction in inference time or energy consumption.

\section{Case Study}
\label{sec:case_study}

\begin{figure}[t]
\begin{center}
\includegraphics[width=0.75\columnwidth]{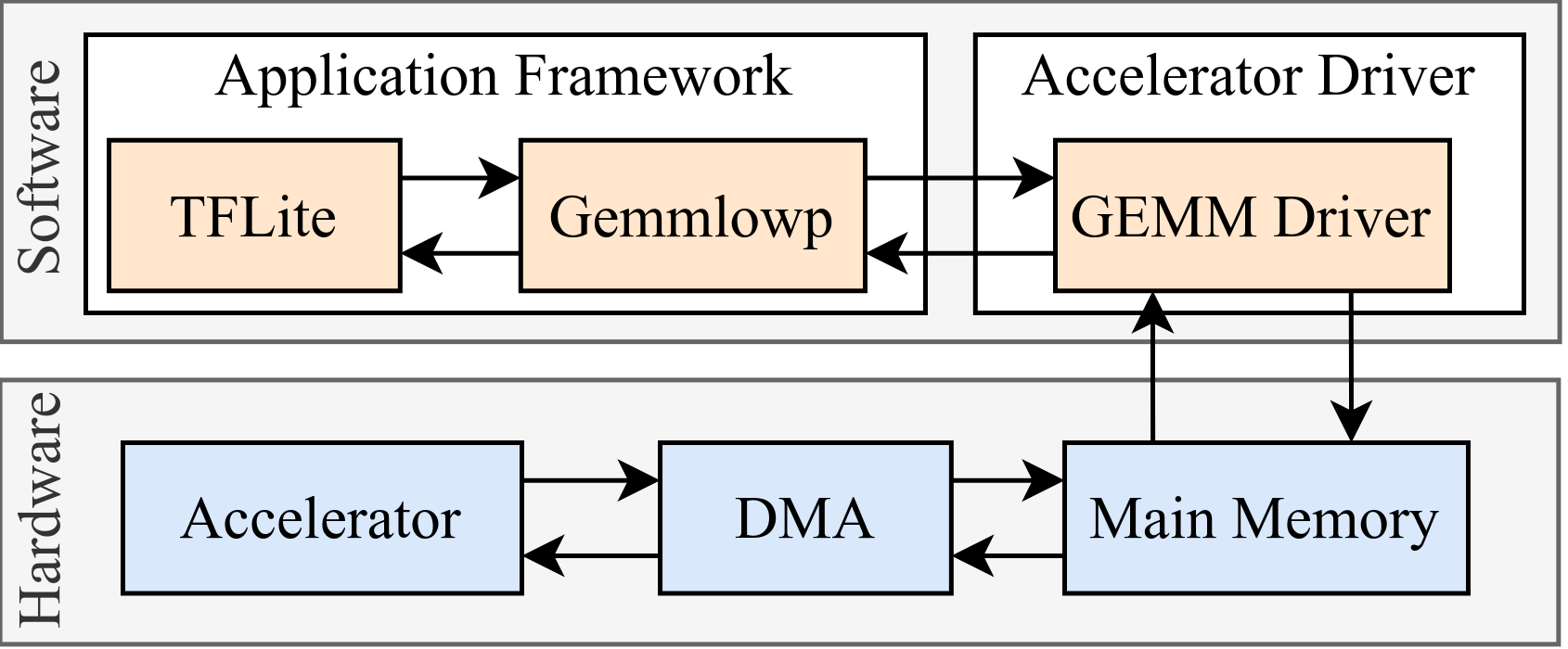}
\caption{\label{fig:runtime_model} Runtime model of our case study: TFLite GEMM convolution using hardware acceleration.}
\end{center}
\end{figure}

To demonstrate the value of the SECDA methodology, we design and implement two different FPGA-based accelerators for DNN inference, a Vector MAC (VM) based-design and a Systolic Array (SA) based-design. The \emph{Application Framework} chosen is TFLite, a popular DNN inference framework for resource-constrained edge devices such as our target device, the PYNQ-Z1 board. We accelerate the convolutional layers, which in TFLite are implemented using the \emph{GEMM convolution} algorithm. Thus, we develop the custom accelerators and their respective drivers to reduce the inference time of the model. Our accelerators use 8-bit quantized DNN models, a popular machine learning approach that can reduce the inference time with a low accuracy penalty~\cite{zhou2017b}. Figure~\ref{fig:runtime_model} shows the execution flow when performing DNN inference using a GEMM accelerator. 
Our accelerator offloading is integrated inside the TFLite runtime through TFLite source code modifications.
We describe the design workflow used throughout the case study and provide details of the designs in the following sections.

\subsection{SECDA Instantiation}

\subsubsection{Initialization}

The initialization step is essential to identify the target workload and to achieve integration within the \emph{Application Framework}. This step varies depending on the goal of the designer (i.e., the target application framework and workload). Hence, it is difficult to automate this step without limiting the capabilities of SECDA as a methodology to work with new application frameworks and workloads. 

Within this case study, SECDA is initialized by integrating end-to-end simulation with TFLite, thus establishing the foundation of our co-design/co-verification environment. 
The initial stage is to identify where in TFLite to intercept GEMM calls, in our case the \emph{Gemmlowp} library, to offload expensive computations to the target accelerator and create an initial implementation to fetch and return its data. 
We use a native C\texttt{++} implementation of the GEMM function, which evolves with the addition of SystemC hardware definitions as we move from the initial stage. Once initialized, we can reuse the development hooks, such as SystemC simulation or function calls, that we have created for the target application framework. 
We reuse the SECDA integrated TFLite codebase from the first design (VM) when developing the second design (SA) to enable a much faster development process.

\subsubsection{SystemC Simulation co-design/co-verification}

After adding simple SystemC constructs to our GEMM accelerator module, we develop hardware components in SystemC to replace the initial implementation. Developing the testbench discussed in Section \ref{subsec:sysc_sim}, we implement hardware components for computing the GEMM function (e.g., weight buffers, multipliers). 
Using the testbench and the end-to-end simulation environment, we go through several iterations where we fine-tune our accelerator components.
For example, reducing the number of clock cycles, or changing the behavior of the \emph{Accelerator Driver} which improves data reshaping.

\subsubsection{Design Loop} 

When we have an accelerator design that has no major bottlenecks in simulation in terms of clock cycles and imposes efficient resource utilization of the target device, we use \emph{Hardware Synthesis} to map it onto the PYNQ-Z1's FPGA. A key strength of SECDA is that benchmarking on real hardware uses the same \emph{Application Framework} (TFLite) and \emph{Application Driver} as the simulated version. The following sections describe the two accelerators designs and their drivers.


\begin{figure}[t]
\begin{center}
\includegraphics[width=0.85\columnwidth]{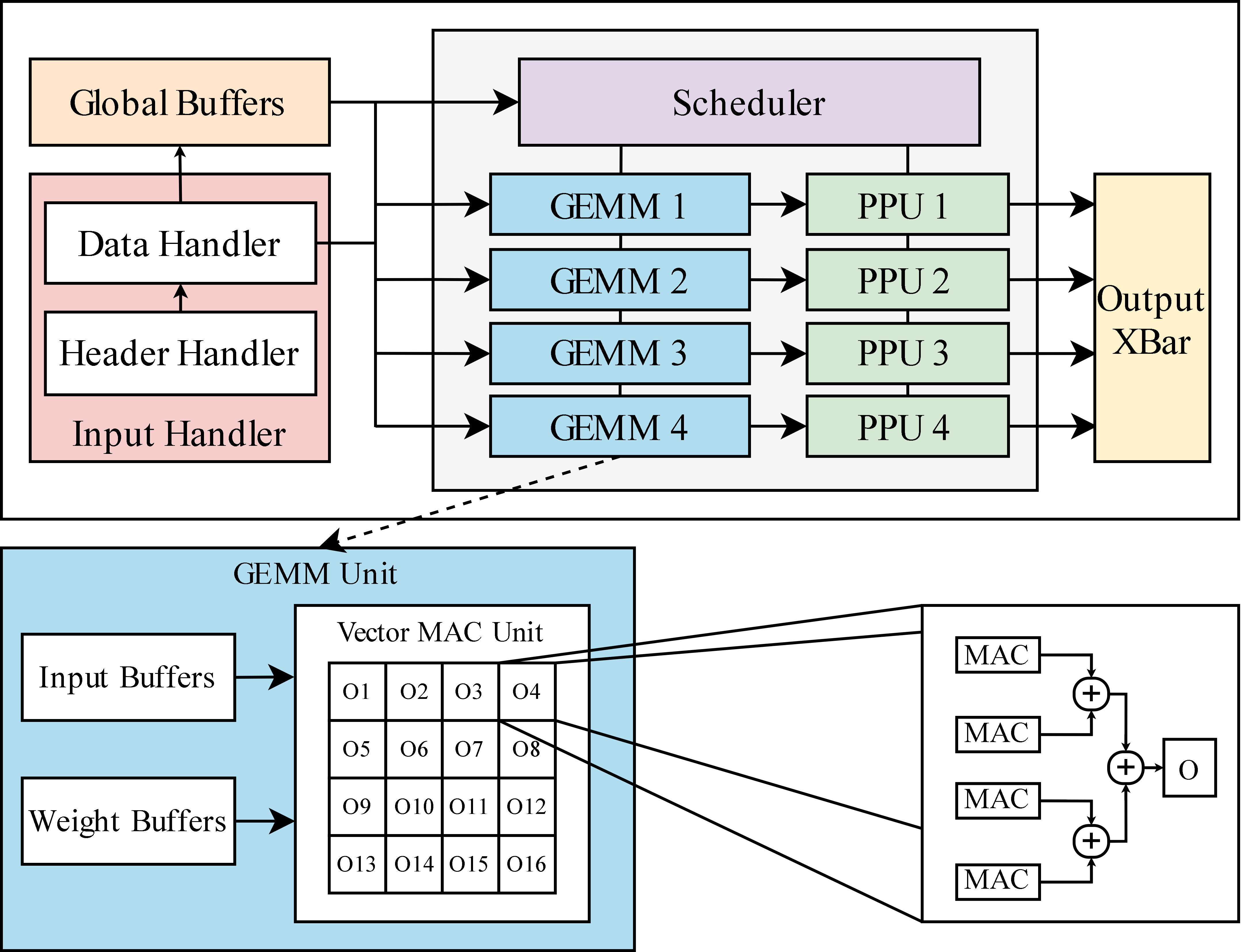}
\caption{\label{fig:full_acc_design}Accelerator design, featuring four GEMM Units.}
\end{center}
\end{figure}

\subsection{GEMM Accelerator Driver}
\label{subsec:gemm_driver}

The software \emph{GEMM Driver} is co-designed with the hardware accelerator and connects to the \emph{Application Framework} TFLite. It intercepts GEMM calls within the \emph{Gemmlowp} library, as shown in Figure~\ref{fig:runtime_model}. 
The GEMM driver handles the execution of convolutional layers utilizing the accelerator. 
It receives both weight and input data from TFLite, and reshapes them to our chosen accelerator data format.

This data format was co-designed with the accelerator, such that 
\begin {enumerate*} [label=\itshape\roman*\upshape)]
\item CPU-side data preparation leverages vectorized loads to reduce transformation overheads; 
\item data is partitioned across multiple memory-mapped buffers, hence can be sent concurrently over the DMA interface, as shown in Figure~\ref{fig:runtime_model};  
\item data in each partition is organized such that it can be distributed efficiently inside the accelerator.
\end {enumerate*}

Once the data is reshaped, the driver is responsible for sending it to the accelerator, and for collecting and storing the output data. We pipelined the execution of the operations within the GEMM driver across multiple batches of GEMM operations within each layer to ensure that the CPU is not idle while the accelerator is processing inputs. In later design iterations, we found that the bottleneck was no longer the GEMM operations. Hence we moved software-side post-processing steps (see Section \ref{subsubsec:ppu}) to the accelerator, with the GEMM driver managing the new functionality. 
Note that the key difference between drivers for VM and SA designs is the handling of output data, as the output layouts differ.


\begin{figure}[t]
\begin{center}
\includegraphics[width=0.85\columnwidth]{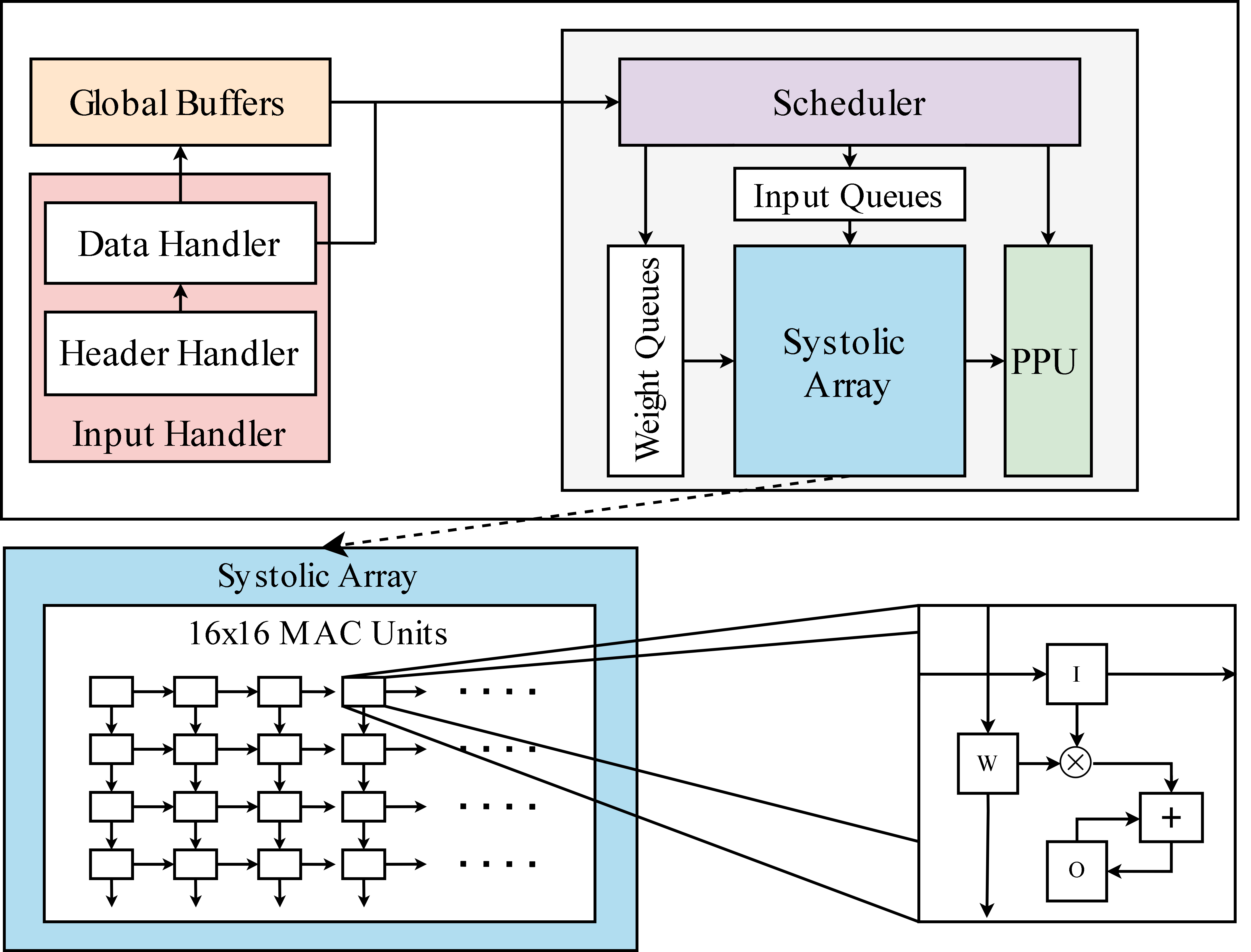}
\caption{\label{fig:sys_array} Accelerator design, featuring a 16x16 Systolic Array.}
\end{center}
\end{figure}

\subsection{GEMM Accelerator Designs}
\label{subsec:acc_design}

Both VM and SA designs follow an output-stationary dataflow approach~\cite{kwon2019}, which was chosen to remove the need to store many intermediate results on valuable on-chip memory, or incur costs associated with storing them off-chip.

\subsubsection{Vector Mac Design (VM)} 

Figure~\ref{fig:full_acc_design} shows an overview of the VM accelerator design that consists of four SIMD-style compute units, which we refer to as GEMM units. We are limited to four GEMM units by the resource constraints of the target device. Each GEMM unit broadcasts sets of weights and inputs them to its internal MAC units to produce $4\times4$ output result tiles. Each output value is calculated using a set of four MAC units, with the intermediate results reduced to the final output value through an adder tree.

\subsubsection{Systolic Array Design (SA)} 

Figure~\ref{fig:sys_array} shows an overview of the SA accelerator design. The design contains a single computation unit constructed as a $16\times16$ MAC-based systolic array, where each MAC unit accumulates towards a single output value. MAC units work by reading and storing the input and weight values of the neighboring MAC units into their own registers. Hence, the systolic array moves weight and input values vertically and horizontally, respectively, once at the start of each step. 
The inputs and weights for the starting row and column of the MAC units are read from a set of 32 data queues which are filled by the scheduler.



\subsection{Accelerator Components}
\label{subsec:components}

Our designs are constructed with basic components, developed and tested both individually in the SystemC testbench, and together in end-to-end simulation. Both designs contain similar components, though their behavior and connections vary.
Adapting, reusing, and recomposing these components for new designs is a valuable feature of any hardware design methodology, especially in DNNs where a given design may lose relevance quickly due to novel DNN workloads emerging. Below is a brief description of the major components.

\subsubsection{The Input Handler}\label{subsubsec:input_handler} receives all data sent by the GEMM Driver from main memory via DMA, as shown in Figure~\ref{fig:runtime_model}. Metadata added by the driver is used to direct the incoming data to the appropriate accelerator buffers. The arrangement of the buffers varies between both designs. The VM design makes use of local buffers within each GEMM unit to store all input values and the active tile of weight data, with the global buffers used for storing all weights tiles; the SA design only uses global buffers for both input and weight data.  

\subsubsection{The Scheduler}\label{subsubsec:scheduler} orchestrates computations which occur within the processing units of each design. For the VM design, the \emph{Scheduler} assigns work to each GEMM unit, broadcasting weight data tiles to all GEMM units, and ensuring maximum weight data tile reuse to minimize redundant loads. For the SA design, the \emph{Scheduler} feeds input and weight data to the corresponding data queues, which feed the outer MAC units within the array. 

\subsubsection{Post Processing Unit (PPU)}\label{subsubsec:ppu} receives \texttt{uint32} output tiles from their adjacent processing unit, and applies the post-processing pipeline to obtain the quantized \texttt{uint8} result tiles. Originally performed on the CPU-side, this data size reduction enabled us to reduce output data transfer costs by $4\times$ at the cost of additional resource usage. Additionally, the PPU performs all other functionality provided by \emph{Gemmlowp}'s ``unpacking'' function, including bias addition, scaling, and applying the activation function. For the VM design, there are multiple smaller PPUs which process the output from each GEMM Unit. The PPU outputs are combined later by the \emph{Output Crossbar}. In comparison, the SA design contains a single PPU which processes all the $16\times16$ output tiles and sends them back to main memory.

\subsubsection{Output Crossbar}\label{subsubsec:xbar} used to collect the output tiles from all PPUs (only VM design). It rearranges the tiles such that the results are sent back to main memory in the desired order.



\subsection{Accelerator Design Improvements}
\label{subsec:design_iterations}

Our SECDA methodology enables fast and iterative development of DNN accelerator designs. Here we discuss the major design improvements we made to optimize the end-to-end performance for both designs.

\subsubsection{Improved Data Distribution \& Bandwidth Utilization}

During the design process for the VM accelerator, in simulation, we observed a lower BRAM bandwidth utilization than expected. To address the low BRAM utilization, we added extra functionality to the \emph{Input Handler} to distribute the incoming input and weight data across multiple BRAMs, increasing the number of data accesses possible per cycle. 

The synthesis of our first VM design consisting of four GEMM units, highlighted a data transfer bottleneck between off-chip and on-chip memory that was not modeled within the simulation. We alleviated this bottleneck by ensuring that we leveraged all of the high-performance AXI data links available on the PYNQ-Z1 board. From this change, we used end-to-end simulation to quickly redesign the accelerator and the accelerator driver to leverage the improved data links, significantly reducing data transfers times.

For the SA design, allocating $32$ data queues to feed the outer MAC units of the systolic array and enabling the \emph{Scheduler} to fill the data queues in parallel with the processing of the systolic array minimized MAC unit idle time within the SA accelerator due to unavailability of data.


\subsubsection{Scheduling \& Post Processing}

For the VM design, the simulation highlighted a slowdown that occurred within each GEMM unit when reading the weight tiles into the local buffers. To address this slowdown we added the \emph{Scheduler Unit}, which improved the ordering of computations, reducing the number of reads from global weight buffers by $4\times$. 

Through \emph{Hardware Execution} we obtained a breakdown of the inference time, which indicated that post-processing performed within the \emph{Gemmlowp} library was the new bottleneck. Hence, we enhanced the capabilities of the accelerators by implementing post-processing within them. By adding the PPU, we obtained $1.5\times$ and $1.3\times$ speedup on single and dual-thread inference, respectively, when compared to previous VM designs without it. 
To move more functionality to the accelerator, we adapted the GEMM driver to receive quantized 8-bit results produced by the post-processing quantization, as opposed to the 32-bit results which are generated by the GEMM operations, reducing output data transfer costs by $4\times$.

\subsubsection{Varying Systolic Array Sizes}

The SA design was prototyped varying the dimensions of the array. We explored $4\times4$, $8\times8$ and $16\times16$ designs, evaluating tradeoffs obtained by varying the output tile sizes and resource utilization. We found in simulation that the $4\times4$ design lacked the compute power for the accelerator to improve against CPU-based GEMM. The $8\times8$ design outperformed the CPU baseline, though left much of the PYNQ Z1 FPGA fabric unused. The $16\times16$ design improved performance by 1.7$\times$ across the various models for single thread inference compared to the $8\times8$ design, at the cost of higher resource utilization of the board.

\subsubsection{DNN Specific Design Optimizations}
\label{sssec:model_based_design_opt}
 
With SECDA, we were able to make model specific changes easily to accelerator designs, either in the host driver code or the accelerator design configurations, to improve the performance for a given model.
Due to device constraints, both SA and VM designs cannot be allocated with enough global weight buffer space to fit some larger layers of InceptionV1 and ResNet18 entirely on the accelerator.
With SECDA's ability to quickly simulate the performance and correctness of new designs, we co-designed a weight tiling scheme that was fast to produce on the CPU side and process in the accelerators.
Compared to the previous accelerator designs, this sped up the average inference time for InceptionV1 and ResNet18 by $2\times$ and $2.2\times$, respectively.

Note that some convolutional layers of ResNet18 were still too large to fit into the local buffers of the VM design.
We were able to reconfigure, validate, and synthesize a modified VM design for ResNet18, which trades off global buffer space for local buffer space, enabling native execution of all layer within the accelerator and reducing the inference time by $1.6\times$ over the previous design.

\section{Evaluation}
\label{sec:eval}


\subsection{Experimental Setup}

We evaluate the two accelerator designs in our case study on the PYNQ-Z1 board, which includes an edge FPGA and a dual-core ARM Cortex-A9 CPU. We benchmark four widely used DNN models quantized to 8 bits: MobileNetV1~\cite{8578384}, MobileNetV2~\cite{sandler2019}, InceptionV1~\cite{szegedy2015} and ResNet18\cite{HeCVPR2016}; all defined on the ImageNet dataset \cite{ILSVRC15}. For each DNN model, we evaluate CPU-only inference times in TFLite using $1$ and $2$ CPU threads, taking the median of $100$ runs, to compare against our two accelerator designs. 
Note that non-accelerated TFLite layers use their native C++ implementations and are compiled with the recommended TFLite optimizations for the target platform. 
Conversely, our accelerated layers use custom-designed CPU-side accelerator drivers (see Section \ref{subsec:gemm_driver}) with handwritten optimizations (e.g., vector instructions). 
We gather energy metrics using a COOWOO digital USB power meter.


\subsection{Case Study Results}

Table~\ref{tab:results} shows the breakdown of inference time and energy consumption for the four DNN models under study for a single image using the CPU ($1$ and $2$ threads) and the two accelerator designs (VM and SA). The time is split between convolutional (CONV) layers, which our accelerators target, and all other (Non-CONV) layers which run on the CPU. For the VM accelerator, we observe an average speedup across models of 3$\times$ and 2$\times$ and an average energy saving of 2.7$\times$ and 1.8$\times$ for one and two threads, respectively, in each case when compared to CPU-only inference. Similarly, for the SA accelerator, we observe an average speedup across models of 3.5$\times$ and 2.2$\times$ and an average energy saving of 2.9$\times$ and 1.9$\times$ for one and two threads, respectively, in each case when compared to CPU-only inference.

\begin{table}[t]
\caption{\label{tab:results} Inference time (ms) and energy consumption (J) results for the four DNN models under study when using different number of CPU threads and accelerator designs.}
\centering
\fontsize{6.9}{9}\selectfont
\begin{tabular}{|c|l|c|c||c|c|}
\hline
\textbf{DNN} & \textbf{Hardware setup} & \textbf{CONV} & \textbf{Non-CONV} & \textbf{Overall} & \textbf{Energy} \\ \hline
\parbox[t]{2mm}{\multirow{6}{*}{\rotatebox[origin=c]{90}{MobileNetV1}}} 
                & CPU (1 thr)       & 635 ms & 141 ms & 776 ms & 1.84 J \\\cline{2-6} 
                & CPU (1 thr) + VM & 123 ms & 141 ms & 264 ms & 0.68 J \\\cline{2-6} 
                & CPU (1 thr) + SA & 90 ms & 141 ms & 231 ms & 0.65 J \\\cline{2-6} \cline{2-6} 
                & CPU (2 thr)       & 329 ms & 73 ms & 402 ms & 1.04 J \\ \cline{2-6} 
                & CPU (2 thr) + VM & 105 ms & 73 ms & 178 ms & \textbf{0.43 J} \\ \cline{2-6} 
                & CPU (2 thr) + SA & 86 ms & 73 ms & \textbf{159 ms} & 0.54 J \\ \hline
 \hline
 \parbox[t]{2mm}{\multirow{6}{*}{\rotatebox[origin=c]{90}{MobileNetV2}}} 
                & CPU (1 thr)       & 526 ms & 176 ms & 702 ms & 1.66 J \\\cline{2-6} 
                & CPU (1 thr) + VM & 156 ms & 176 ms & 332 ms & 0.79 J\\\cline{2-6} 
                & CPU (1 thr) + SA & 103 ms & 176 ms & 279 ms & 0.83 J\\\cline{2-6}  \cline{2-6} 
                & CPU (2 thr)       & 277 ms & 95 ms & 372 ms & 1.01 J \\ \cline{2-6}
                & CPU (2 thr) + VM & 128 ms & 95 ms & 223 ms & \textbf{0.61 J} \\ \cline{2-6} 
                & CPU (2 thr) + SA & 97 ms & 95 ms & \textbf{191 ms} & \textbf{0.61 J} \\ \hline
 \hline
 \parbox[t]{2mm}{\multirow{6}{*}{\rotatebox[origin=c]{90}{InceptionV1}}} 
                & CPU (1 thr)       &1416 ms & 117 ms & 1533 ms & 3.60 J \\\cline{2-6} 
                & CPU (1 thr) + VM & 263 ms & 117 ms & 380 ms & 0.97 J \\\cline{2-6} 
                & CPU (1 thr) + SA & 225 ms & 117 ms & \textbf{342 ms} & 1.12 J \\\cline{2-6} \cline{2-6} 
                & CPU (2 thr)       & 736 ms & 117 ms & 853 ms & 2.20 J \\ \cline{2-6} 
                & CPU (2 thr) + VM & 249 ms & 117 ms & 366 ms  & \textbf{0.97 J} \\ \cline{2-6}
                & CPU (2 thr) + SA & 225 ms & 117 ms & \textbf{342 ms}  & 1.12 J \\ \hline
 \hline
 \parbox[t]{2mm}{\multirow{7}{*}{\rotatebox[origin=c]{90}{ResNet18}}}
        & CPU (1 thr)       & 1762 ms & 132 ms & 1894 ms & 5.4 J \\\cline{2-6} 
        & CPU (1 thr) + VM  & 555 ms & 132 ms & 687 ms & 2.12 J \\\cline{2-6} 
        & CPU (1 thr) + SA  & 405 ms & 132 ms & \textbf{537 ms} &  1.76 J \\\cline{2-6} \cline{2-6} 
        & CPU (2 thr)       & 919 ms & 132 ms & 1051 ms & 3.24 J \\ \cline{2-6} 
        & CPU (2 thr) + VM  & 550 ms & 132 ms & 682 ms  & 2.12 J \\ \cline{2-6} 
        & CPU (2 thr) + SA  & 405 ms & 132 ms & \textbf{537 ms}  & 1.76 J \\ \cline{2-6}
        & CPU (2 thr) + VTA & -- & -- & 737 ms  & \textbf{1.51 J} \\ \hline
\end{tabular}

\end{table}

We observe less speedup and energy consumption with dual-thread execution as expected, since the compute capacity of the CPU doubles, while both accelerator designs remain the same.
However, our accelerated runtime using two threads improves inference time since the CPU-side \emph{Accelerator Driver} can leverage threads. While analyzing our designs, we observe that we hit a threshold for performance gains achieved by our hardware designs, with the bottleneck for inference performance shifting to two other areas. Namely,
\begin{enumerate*}[label=(\roman*)]
\item \label{itm:data-prep} CPU-side CONV data preparation and result unpacking; 
\item \label{itm:non-conv} and non-accelerated layers.
\end{enumerate*}
For \ref{itm:data-prep}, breaking down single-threaded CONV time for VM, we observe that only $31\%$ of the time is spent performing off-chip data transfers and the accelerator computations. The CPU-side data preparation and resulting unpacking represent the majority of the CONV time, $69\%$, which highlights the importance of hardware/software co-design to ensure that additional hardware changes cannot further reduce this time. For \ref{itm:non-conv}, in single thread CPU-only inference, Non-CONV layers only represent $14\%$ of the inference time on average. However, by accelerating the CONV layers, the relative importance of Non-CONV layers increases, representing $39\%$ and $46\%$ of single thread inference time for VM and SA, respectively. Comparing our two designs, SA achieves slightly better performance, $16\%$ on average in latency and up to $4\%$ in energy savings. From these observations, we conclude that while the core compute units of VM and SA use different strategies to perform GEMM, we achieve similar end-to-end performance from both designs, due to the shift in the inference performance bottlenecks to the CPU-side.

We also observe that InceptionV1 achieves the best speedup relative to the CPU-only version, with $4\times$ and $2.3\times$ speedup for one and two threads, respectively, for VM, and $4.5\times$ and $2.5\times$, respectively, for SA. Comparing to MobileNetV1 and MobileNetV2, which feature depthwise separable convolutions (meaning that each convolutional layer performs fewer MACs per input), InceptionV1's standard convolutions have greater potential for GEMM acceleration, since the relative cost of its data preparation stage is smaller. Additionally, for InceptionV1 and ResNet18 we observe negligible speedup for multi-threaded execution, relative to the other models due to the larger GEMM operations coupled with our pipelined execution. 
This means that the CPU-side latency, due to data format conversions, is ``hidden'' by the accelerator's computation, and thus results in minimal benefits from two threads.


Note that both accelerator designs could still be further refined. However, the purpose of the case study is to highlight that by using SECDA, we were able to quickly develop and iterate upon viable accelerator designs that significantly improve inference time performance and energy consumption against the CPU-only case. 
Providing a fair experimental comparison of development time between two design methodologies is difficult since measurements could be influenced by the designer's experience and which methodology is evaluated first.
Therefore Equation~\ref{eqn:secda_time} is used to provide estimates of improvements in development times. By replacing synthesis iterations by simulations, we observed a $25\times$ difference between $S_{\mathrm{t}}$ and $C_{\mathrm{t}}$. This suggests that we spent on average $16\times$ less time evaluating end-to-end inference of a given design in simulation for our GEMM accelerators, compared to developing with all evaluation performed on an FPGA.


\subsection{Comparison with state-of-the-art DNN accelerators}

We now validate that our designs are competitive with another state-of-the-art DNN accelerator in terms of inference time, our main design goal.
We compare our designs against VTA, which is supported through the state-of-the-art DNN compiler framework TVM~\cite{tvm}.
We chose it over other accelerator frameworks due to its recent release, support from an active open source community, and its use of 8-bit quantization similar to our designs.
The final row of Table~\ref{tab:results} shows the performance of VTA for ResNet18, taking the median of $100$ runs on the PNYQ Z1 board.
ResNet18 was the only publicly available model which was compatible with both VTA and TFLite.
We refer the reader to the TVM VTA documentation\footnote{\url{https://tvm.apache.org/docs/vta/tutorials/frontend/deploy_classification.html\#sphx-glr-vta-tutorials-frontend-deploy-classification-py}} for details on synthesis and execution --- note that VTA leverages both threads of the CPU.
The results show that the designs developed using the SECDA methodology are competitive with VTA, with our VM design outperforming VTA by $8\%$ in terms of latency, while VTA reports $29\%$ less energy consumption per inference.
Our SA design outperforms VTA by $37\%$ in terms of latency, while VTA has $14\%$ lower energy consumption.
VTA runs more of its layers on the accelerator, which results in fewer off-chip data transfers, therefore achieving a greater energy efficiency than our design.
In terms of our target performance metric, inference time, we have demonstrated that designs produced via SECDA can be competitive with a state-of-the-art accelerator.

\section{Related Work}
\label{sec:related_work}

There are a range of design tools for both DNN accelerators and co-design workflows in other domains.
STONNE \cite{stonne-iiswc2021} provides cycle-accurate simulation for deep learning accelerator designs such as MAERI~\cite{kwon2018} and SIGMA~\cite{qin2020}.
However, it does not integrate full system simulation as SMAUG does, and similar to SMAUG, does not have a direct path to map candidate designs to real hardware.
TFLITE-SOC~\cite{Tflite-soc_2020} features aspects of the SECDA methodology by tightly integrating its \emph{Application Framework} (TFLite) with a SystemC system-level simulation.
SystemC has been used as a part of co-design methodologies in other domains such as cryptographic SoCs~\cite{khalil-hani2008} and image processing~\cite{chong2011}, which demonstrate the streamlined development time advantages of leveraging SystemC.
SECDA focuses the advantages of SystemC on the problem of DNNs, by proposing a methodology which complements the structure of DNNs, such as multiple layers of varying shapes, some of which must be run on the CPU.

As discussed in Section~\ref{sec:motivation}, an optional stage to DNN accelerator design is automatic hardware space exploration for a given DNN model, which we refer to as \emph{accelerator frameworks}.
Examples include VTA~\cite{moreau2019a}, which defines a GEMM unit and high-level task ISA, built on top of the TVM~\cite{tvm} compiler stack, to produce specific designs for a given DNN architecture. VTA can optionally leverage the AutoTVM tuning tool~\cite{chen2018c} for additional design space exploration. DNNBuilder~\cite{zhang2018a} produces DNN specific architectures and permits exploration of alternative quantization schemes, as adopted by other schemes~\cite{wei2017,Guan2017,umuroglu2017}. 
The \emph{accelerator framework} approach can be extended with DNN/FPGA co-design, such that a specific accelerator design is generated in conjunction with a DNN architecture to more efficiently solve the problem task~\cite{hao2019,hao2019a,jiang2020,abdelfattah2020}.
The process of turning an accelerator design into an accelerator framework is possible following the principles of SECDA. However, in our case study we instead focused on two general purpose accelerator designs to demonstrate the core SECDA methodology.

Finally, a general objective for DNN acceleration is to improve the efficiency of MAC operations. Approaches include improving data reuse and different dataflow strategies~\cite{sze2017}, and leveraging sparsity in weights/activations to reduce MACs, supported by efficient hardware designs~\cite{huang2019,lu2017}. SECDA enables shorter development time and efficient exploration of hardware solutions for these opportunities.

\section{Conclusion}
\label{sec:conc}

In this paper we presented SECDA, a hardware/software co-design methodology for efficient DNN accelerator design targeting edge devices with FPGAs. 
SECDA tightly integrates accelerator design with the target application, improving opportunities for co-design of the accelerator driver. 
By simulating the accelerator behavior using SystemC, we reduce development costs by minimizing the number of synthesis iterations, and allow fine-grained benchmarking and co-verification of accelerator components. Further, we can directly map our simulated designs to an FPGA without re-implementation. 
As a case study, we proposed two GEMM-based accelerator designs for optimizing inference of four DNN models using TFLite and a PYNQ-Z1 board. The accelerated models outperform the CPU baseline in all cases. As future work, we plan to automate aspects of SECDA for further reduction in development time, and support other convolution strategies (e.g., Winograd) or DNN classes (e.g., Transformer models). 


\section*{Acknowledgment}
This work was partially supported by the Engineering and Physical Sciences Research Council (grant EP/R513222/1).


\bibliography{references}

\bibliographystyle{IEEEtran}

\end{document}